\documentstyle[11pt,aaspp4]{article}

\begin{document}

\title{Evolution of Multiphase Hot 
Interstellar Medium in Elliptical Galaxies}
\author{Yutaka Fujita\footnote{Present Address : 
Department of Physics, Tokyo Metropolitan University, 
Minami-Ohsawa 1-1, Hachioji, Tokyo 192-03, Japan : 
yfujita@phys.metro-u.ac.jp}}
\affil{Graduate School of Human and Environmental Studies, \\
Kyoto University, Kyoto 606-01, Japan}
\affil{Yukawa Institute for Theoretical Physics, \\
Kyoto University, Kyoto 606-01, Japan}
\authoraddr{Kyoto 606-01, Japan}
\authoremail{fujita@yukawa.kyoto-u.ac.jp}
\author{Junji Fukumoto}
\affil{Nihon Silicon Graphics Cray K.K., Cuore Bldg., 9th Floor, 12-25,\\
        Hiroshiba-cho, Suita-si, Osaka 564, Japan}
\author{Katsuya Okoshi}
\affil{Department of Earth and Space Science,Faculty of Science,\\
Osaka University,Machikaneyama-cho, Toyonaka, Osaka 560, Japan}
\authoremail{okoshi@vega.ess.sci.osaka-u.ac.jp}

\begin{center}
{\em To appear in The Astrophysical Journal}
\end{center}

\begin{abstract}
We present the results of a variety of simulations concerning the evolution 
of multiphase (inhomogeneous) hot interstellar medium (ISM) in elliptical 
galaxies. We assume the gases ejected from stars do not mix {\em globally} 
with the circumferential gas. The ejected gas components evolve separately 
according to their birth time, position, and origin. We consider cases where 
supernova remnants (SNRs) mix with local ISM. The components with high metal 
abundance and/or high density cool and drop out of the hot ISM gas faster 
than the other components because of their high metal abundance and/or 
density. This makes the average metal abundance of the hot ISM low. 
Furthermore, since the metal abundance of mass-loss gas decreases with 
radius, gas inflow from outer region makes the average metal abundance of 
the hot ISM smaller than that of mass-loss gas in the inner region. As gas 
ejection rate of stellar system decreases, mass fraction of mass-loss gas 
ejected at outer region increases in a galaxy. If the mixing of SNRs is 
ineffective, our model predicts that observed [Si/Fe] and [Mg/Fe] should 
decrease towards the galactic center because of strong iron emission by SNRs.
In the outer region, where the cooling of time of the ISM is long, the 
selective cooling is ineffective and most of gas components remain hot. Thus,
the metal abundance of the ISM in this region directly reflects that of the 
gas ejected from stars. Our model shows that supernovae are not effective 
heating sources in the inner region of elliptical galaxies, because most of 
the energy released by them radiates. Therefore, cooling flow is established 
even if the supernova rate is high. Mixing of SNRs with ambient ISM makes the 
energy transfer between supernova explosion and ambient ISM more effective.
\end{abstract}


\section{INTRODUCTION}
\label{sec-1}
\indent

Observations by X-ray satellites have shown that the X-ray properties of
bright elliptical galaxies can be explained by thermal emission of 
hot interstellar medium (ISM).
The estimated temperatures are 
around 1 keV and X-ray luminosities are typically around $10^{40}$ erg. 
Theoretical arguments indicate that the ISM is inhomogeneous;
Mathews \markcite{m1990} 
(1990) estimated that
the $\sim 1 M_{\sun}$ of metal ejected by each supernova 
event into the ISM 
is trapped locally within the hot bubbles.
Since in elliptical galaxies,
there is no 
overlapping of expanding supernova remnants after galactic wind period 
(Fujita, Fukumoto, \& Okoshi \markcite{ffo1996}1996 ; hereafter Paper I), 
it is expected that this 
inhomogeneity persists for a long time.
The recent observation of {\em ROSAT} 
supports this idea by showing existence of 
components with several temperatures in elliptical
galaxies. (Kim \& Fabbiano \markcite{kf1995}1995)

Based on these arguments, we 
studied the evolution of the multiphase (inhomogeneous) 
ISM in Paper I. We showed that 
the gas components with high metal abundance and high density
cool and drop out of the hot ISM faster than other components. 
As gas ejection rate of stellar system decreases, gas components with
low metal abundance dominate in a galaxy. 
As a result, the average metal abundance which we predict 
is lower than that  
predicted by the previous homogeneous ISM models 
(e.g. Arimoto \& Yoshii \markcite{ay1987}1987 ; 
Matteucci \& Tornamb\'{e} \markcite{mt1987}1987 ; 
Loewenstein \& Mathews \markcite{lm1991}1991 ; 
Renzini et al. \markcite{rcdp1993}1993 ; 
Mihara \& Takahara \markcite{mg1995}1994 ; 
Matteucci \& Gibson \markcite{mg1995}1995 ; 
Fukumoto \& Ikeuchi \markcite{fi1996}1996). The low metal 
abundance is consistent with the observations. 
(Awaki et al. \markcite{amt1994}1994 ;  Loewenstein et al.
\markcite{lmt1994}1994 ; Matsushita 
et al. \markcite{mma1994}1994 ; Mushotzky et al. \markcite
{mla1994}1994 ; 
Kim \& Fabbiano \markcite{kf1995}1995 ; Davis \& White
\markcite{dw1996}1996 ; Matsumoto et
al. \markcite{m1996}1996).  
We also predicted in Paper I that the heating by supernovae is not 
effective because the gas components originated from supernovae
remnants have high metal abundance, and their energy radiates before
transfered to the circumferential ISM.

However, since our model was a simple 
one-zone model, it could not predict spatial 
variation in the ISM, which is recently observed by 
{\em ASCA} (Mushotzky et al. \markcite
{mla1994}1994)
and {\em ROSAT} (Kim \& Fabbiano \markcite{kf1995}1995 ; Rangarajan et
al. \markcite{r1995}1995 ; Irwin \& Sarazin \markcite{is1996}1996). 
In this paper, we investigate the radial evolution of the ISM 
assuming that the gases ejected from stars do not mix with
the ambient ISM globally.

The paper is organized as follows. In \S2 we
describe our method for simulating the evolution 
of the multiphase hot 
ISM in elliptical galaxies. 
Then, in \S3, we solve these equations numerically for some
typical models. 
Our results are summarized in \S4. 

\section{ASSUMPTION AND BASIC EQUATIONS}
\label{sec-2}
\indent

We consider the evolution of the hot ISM after galactic wind period.
We assume for simplicity that an elliptical galaxy is  
spherically symmetric. 
The model galaxies we used here are the same as those of 
Loewenstein \& Mathews
(1987). 
The stellar density distribution is
\begin{equation}
\label{star-eq}
\rho_{\star}(R)=\rho_{0 \star}[1+(R/R_{a\star})^{2}]^{-3/2} \:
\end{equation}
for $R<R_{t}$, where $R$ is the distance form
the center of the galaxy and $R_{t}$ is the tidal radius.  
The density distribution of dark halo is
\begin{equation}
\rho_{h}(R)=\rho_{0 h}[1+(R/R_{a h})^{2}]^{-1} \:
\end{equation}
for $R<R_{t}$. The subscripts $a$ and 0 refer to core and central
properties. Both densities are assumed to vanish for $R>R_{t}$.
The properties of the model galaxies are shown in Table 1.
The stellar velocity dispersion profile, $\sigma_{\star}(R)$,
is derived by equation (23) of Sarazin \& White \markcite{sw1987}
(1987), which is valid for an
isotropic velocity dispersion. 

The ISM of the 
galaxy is divided into finite number
zones, $0<R_{1}(t)<...<R_{j}(t)<...<R_{m}(t)$, 
Each zone moves with the ISM.
The zone, $R_{j-1}<R<R_{j}$, is called the $j$th zone.
When $R_{j-1}>R_{t}$, the $j$th zone is deleted. Furthermore, when
$R_{j}<0.12$ kpc, the $j$th zone is also deleted to avoid too many 
time-steps. 
Each zone is assumed to be uniform on
average, although it contains many
gas components (hereafter called ``phases''). 
For simplicity, 
we assume that the phases do not globally 
mix with ambient gas and comove with 
the zones which they belong to.  
These assumptions are valid as long as magnetic field suppresses
relative 
motion of the phases (Hattori, Yoshida, \&
Habe \markcite{hyh1995}1995). We will discuss the validity in
\S\ref{sec-mag}. 
The phases are classified 
by their origin, birth time, and zone which they belong to.
We consider the three types of the origin, that is, 
gas ejected through stellar wind (mass-loss
gas), shell and cavity of supernova remnants (SNRs).
The phases originated from mass-loss gas are called ``mass-loss
phases'' indicated by the index $ML$.
Since the cavity of a SNR is buoyant, it floats and breaks part of the shell,
and mixes with the ambient ISM {\em
locally}. These partially broken shell and mixed cavity evolve as
separate phases. 
They are called ``bored shell phases (BSPs)'' and ``mixed cavity phases 
(MCPs)'' indicated by the indices 
$bs$ and $mc$, respectively.
We divide time into a finite number of steps for the $j$th zone, 
$0<t_{0,j}<t_{1,j}<...<t_{i,j}<...<t_{n,j}$ (see equation(\ref{step})), where 
$t_{0,j} (=t_{0})$ and $t_{n,j} (=t_{f})$ 
are the start and end time of the calculation, respectively.
The both times are independent of the zones. 
We define $t=t_{0}(=0.5 \rm Gyr)$ 
as the time when the galactic wind stops.
In each time-step,
one phase is born for each type.
The phase of type $\alpha$ ($ML$, $bs$, or $mc$) 
which is born in $t_{i-1,j} < t < t_{i,j}$ and in the $j$th zone
is called the $(i,j,\alpha)$-th phase and 
has its own temperature $T^{(i,j,\alpha)}$, 
density $\rho^{(i,j,\alpha)}$, metal abundance 
$Z^{(i,j,\alpha)}$, and
mass $M^{(i,j,\alpha)}$.
Each phase radiates its thermal energy and evolves. 
We assume that energy
transfer between the phases is worked by pressure. Therefore, 
the phases are {\em not} independent each other. We ignore
thermal conduction for simplicity.
The phases are assumed to be in pressure equilibrium 
because sound crossing time of an elliptical galaxy is far 
shorter than its age. 
Exceptionally, the phases whose
temperatures become below $T_{\rm crit} (= 10^{5}\rm K)$ are not
considered to be in the pressure equilibrium 
because the condition of the pressure equilibrium will 
broken down for the phases due to the high cooling rate. 
They are assumed to 
cool immediately and drop out of the hot ISM. 
The gas 
left after galactic wind in the $j$th zone 
is also considered as a phase called
``zero-phase''or $(0,j,0)$-th phase. 

In \S\ref{sec-fm}, we describe the formation of the phases in the
$j$th zone at  
$t=t_{i,j}\:(1 \leq i \leq n, 1 \leq j \leq m)$ unless otherwise mentioned. 
We will often omit the subscripts $i$ and $j$ in \S\ref{sec-fm}.
In \S\ref{sec-beq}, we describe the 
evolution of the phases and the galaxy.

\subsection{FORMATION OF EACH PHASE}
\label{sec-fm}

\subsubsection{Mass-Loss Phases}
\label{sec-ml}
\indent

We assume that the temperature of the gas ejected 
by the stellar wind immediately becomes equal to 
the stellar temperature of the galaxy $T{\star}(R)=
[\sigma_{\star}(R)]^{2}\mu m_{H}
/k_{B}$, where $\mu$ is the mean molecular weight ($=0.6$), 
$m_{\rm H}$ is 
the mass of the hydrogen atom, and $k_{\rm B}$ is Boltzmann constant.  

Under this assumption, 
the initial temperature and density of the mass-loss 
phase are then given by 
\begin{equation}
T^{(i,j,ML)}(t_{i,j}) =  T_{\star}(R_{j}) \:,
\end{equation}
\begin{equation}
\rho^{(i,j,ML)}(t_{i,j}) = \frac{\mu m_{\rm H} P^{(j)}(t_{i,j})}
{k_{\rm B}T_{\star}(R_{j})}
  \:,
\end{equation}
where $P^{(j)}(t)$ is the average pressure of 
the ISM in the $j$th zone. The pressure 
$P^{(j)}(t)$ is obtained by solving the evolution equations of the galaxy 
described in \S\ref{sec-beq}.

The gas and iron 
mass of the mass-loss phase at its birth time,  
$M^{(i,j,ML)}(t_{i,j})$  
and $M_{\rm Fe}^{(i,j,ML)}(t_{i,j})$, respectively, are given by 
\begin{equation}
M^{(i,j,ML)}(t_{i,j})=\int_{t_{i-1,j}}^{t_{i,j}}L_{\star}^{(j,ML)}(t)dt \: ,
\end{equation}
\begin{equation}
M_{\rm Fe}^{(i,j,ML)}(t_{i,j})=\int_{t_{i-1,j}}^{t_{i,j}}
L_{\rm Fe}^{(j,ML)}(t)dt \:,
\end{equation}
where 
$L_{\star}^{(j,ML)}$ and $L_{\rm Fe}^{(j,ML)}$ 
are the gas and iron 
mass ejection rates from stars in the $j$th zone, respectively. 
Since the time-scale of star formation in an elliptical galaxy, which is
typically $10^{7-8}$ yr, is short enough 
compared with the galaxy age 
(e.g. Arimoto \& Yoshii\markcite{ay1987} 1987), we assume
that the stellar system of the galaxy formed at $t=0$ simultaneously. 
Thus, $L_{\star}^{(j,ML)}$ and $L_{\rm Fe}^{(j,ML)}$ are given by 
\begin{equation}
\label{star-ml}
L_{\star}^{(j,ML)}(t) = f \left|\frac{d m(\tau)}{d\tau}
\right|_{\tau=t} \phi(m(\tau=t)) 
M_{\star}^{(j)}(t)  \:,
\end{equation}
\begin{equation}
\label{fe-ml}
L_{\rm Fe}^{(j,ML)}(t) = Z_{\rm ML} f \left|
\frac{d m(\tau)}{d\tau}\right|_{\tau=t}
\phi(m(\tau=t))  
M_{\star}^{(j)}(t)  \:,
\end{equation}
where $\phi(m)$ and $Z_{\rm ML}$ are the initial mass
function (IMF) and   
the iron abundance of the mass-loss gas, respectively. 
The power of the IMF is taken to be 1.35. 
Total mass of the stars
in the $j$th zone, $M_{\star}^{(j)}(t)$, is given by
\begin{equation}
M_{\star}^{(j)}(t)=\int_{R_{j-1}(t)}^{R_{j}(t)}4\pi 
R^{2}\rho_{\star}(R)dR \:.
\end{equation}
The relation between the mass of a star $m(\tau)$, and its lifetime  
$\tau$, is given by 
\begin{equation}
\label{turn}
\log_{10}m(\tau) = 
1.983 - 1.054\sqrt{\log_{10}\tau + 2.52}
\:,
\end{equation}
where the mass is in units of solar mass ($M_{\sun}$) and the lifetime is in
units of Gyr.
(Larson\markcite{l1974} 1974). 
Because we consider the evolution of the hot ISM after galactic wind 
stops ($t> t_{0}=0.5 $Gyr), 
this equation implies that we do not have to consider stars with mass
larger than $2.85 M_{\sun}$ in our
calculation period. 
We assume that
stars with mass in the range of $0.1 - 2.85 M_{\sun}$ 
lose their masses by stellar winds ; 
we simply assume that the mass loss occurs instantaneously 
at the end of the life given by the fraction $f$ of the
initial mass of a star with $m$, 
\begin{equation}
\label{frac-ml}
f = \left\{\begin{array}{ll}
                     0 & \mbox{for $m \leq 0.7$}\:,\\
0.42 m & \mbox{for
$0.7 < m \leq 1.0$}\:,\\ 
0.8 - 0.43/m  & \mbox{for
$1.0 < m \leq 2.85$}\:,\\
 		 \end{array}	 \right. 
\end{equation}
(K\"{o}ppen \& Arimoto\markcite{ka1991} 
1991).

Although the metal abundance distribution of the stars or 
mass-loss gas is not
well-known beyond the effective radius, it is assumed to be
\begin{equation}
\label{eq-zml}
Z_{\rm ML}(R)=Z_{\rm ML}(R_{e})(R/R_{e})^{-1/2} \:
\end{equation}
(Arimoto et al. \markcite{m1997}
1997), where $R_{e}$ is the effective radius and we assume that
$R_{e}=12R_{a \star}$.

The density $\rho^{(i,j,ML)}(t_{i,j})$, the temperature 
$T^{(i,j,ML)}(t_{i,j})$, the mass $M^{(i,j,ML)}(t_{i,j})$, 
and the iron mass $M_{\rm
Fe}^{(i,j,ML)}(t_{i,j})$ determined by above equations give 
the initial conditions of the
evolution equations of the phases (see \S\ref{sec-beq}).

\subsubsection{Bored Shell and Mixed Cavity Phases}
\label{sec-sn}
\indent

Next, we describe the formation of the shell and cavity phases. 
We consider only Type Ia supernovae (SN Ia) 
because we are concerned with the 
evolution of the hot ISM after galactic wind stops. 

In Paper I, we assumed that the SNRs consist of two parts,
outer shell and inner cavity region, and
that they evolve as separate phases. We called 
the former ``shell phase'' and the latter ``cavity
phase'', denoted by the indices $s$ and $c$, respectively.
When the distance from the center of the SNR is $r$ and 
the SNR radius is $r_{\rm s}$, which is referred as ``shock front
radius'' in Paper I,   
the inner cavity region and the outer shell region correspond to 
$r<(1-k)r_{\rm s}$ and $(1-k)r_{\rm s} < r < r_{\rm s}$, respectively, 
where $k$ is representing the width of the shell.  
The SNR radius is given by
\begin{eqnarray}
\label{rs}
r_{\rm s}& = & \left[\frac{8}{25(\gamma+1)}\right]^{1/3}
         (P^{(j)})^{-1/3} (2.02 E_{\rm SN})^{1/3} \\
	&\sim& 70 \left(\frac{P^{(j)}}{10^{5} \rm cm^{-3} K}\right)^{-1/3}
        \left(\frac{E_{\rm SN}}{10^{51} \rm erg}\right)^{1/3}  \rm pc \:,
\end{eqnarray}
where $\gamma(=5/3)$ is adiabatic constant, 
and $E_{SN}(=10^{51}$erg) 
is the energy released from 
a supernova (Paper I).

The density, temperature, and mass of the 
shell and cavity region before the cavity floats by buoyant force are
\begin{equation}
\label{eq-rhos}
\rho_{s}=\int_{(1-k)r_{\rm s}}^{r_{\rm s}}4\pi \rho_{\rm Sd}(r) r^{2}dr/
         \int_{(1-k)r_{\rm s}}^{r_{\rm s}}4\pi r^{2}dr \:,
\end{equation}
\begin{equation}
\rho_{\rm c}= \frac{3m_{\rm c}}{4\pi (1-k)^{3} r_{\rm s}^{3}} \:,
\end{equation}
\begin{equation}
T_{\rm s}=\frac{2}{3}\frac{\mu m_{\rm H} U_{\rm s}}{k_{\rm B}}
+\hat{T}^{(j)}\:,
\end{equation}
\begin{equation}
\label{tc}
T_{\rm c}=\frac{2}{3}\frac{\mu m_{\rm H} U_{\rm c}}{k_{\rm B}}
+\hat{T}^{(j)}\:,
\end{equation}
\begin{equation}
m_{\rm s}=\rho_{\rm s}\int_{(1-k)r_{\rm s}}^{r_{\rm s}}4\pi r^{2}dr \:,
\end{equation}
\begin{equation}
m_{\rm c}=m_{\rm SNR}-m_{\rm s} \:,
\end{equation}
respectively, where $\rho_{\rm Sd}(r)$ is the density
distribution of the Sedov solution, $U$ is the specific thermal
energy, $\hat{T}^{(j)}$ 
is the average temperature 
of the ISM in the $j$th zone, and $m_{\rm SNR}$ is the mass 
inside the SNR radius. Their derivations are described in
Paper I. Note that there is no physical 
meaning in using Sedov solution ; it is 
only an approximation of Fig.2 - 4 of Mathews \markcite{m1990}(1990)
as we discussed in Paper I.

In Paper I, we ignored floating of the hot cavity 
region by buoyant force and subsequent mixing by Rayleigh-Taylor 
instability (buoyant mixing) for simplicity,  
although Mathews \markcite{m1990}(1990) indicated that it is effective.
In this paper, we consider the effect as the followings. 

The time-scale for the cavity region to move by one
diameter and to mix into the local ISM is $\lesssim 10^{6-7}$ yr (Mathews
\markcite{m1990}1990). Since it is less than the time-scale of the
evolution of elliptical galaxies, we assume that the cavity region
moves by one diameter and mixes into the local ISM immediately 
after its birth. Figure 1 gives an outline. The region,
$A+B+C$, that is, the local ISM, part of the shell region, and
the cavity region, is assumed to mix uniformly. Hereafter, we refer to the
region $A+B+C$ as the mixed cavity region, and refer to the region 
$D$ as the bored shell region, 
denoted by the indices $mc$ and $bs$, respectively.
They evolve as separate phases. 
We call the former ``mixed cavity
phases (MCPs)'' and the latter ``bored shell phases (BSPs)''. 

The density, temperature, and mass of the bored shell region and mixed 
cavity region are
\begin{equation}
\label{eq-rhobs1}
\rho_{bs}=\rho_{s} \:,
\end{equation}
\begin{equation}
\label{eq-rhomc1}
\rho_{mc}=\frac{m_{c}+\rho_{s}V_{\rm B}+\rho^{(j)} V_{\rm A}}
{V_{\rm A}+V_{\rm B}+V_{\rm C}} \:,
\end{equation}
\begin{equation}
\label{eq-tbs1}
T_{bs}= T_{s} \:,
\end{equation}
\begin{equation}
\label{eq-tmc1}
T_{\rm mc}=\frac{2}{3}\frac{\mu m_{\rm H}}{k_{\rm B}}\frac{E_{\rm
SN}-m_{bs}U_{s}}{m_{mc}}+\hat{T}^{(j)}\:,
\end{equation}
\begin{equation}
\label{eq-mbs}
m_{bs} = \rho_{s}V_{\rm B} \:,
\end{equation}
\begin{equation}
\label{eq-mmc}
m_{mc} = m_{c}+\rho_{s}V_{\rm B}+\rho^{(j)} V_{\rm A} \:,
\end{equation}
where $V_{\rm A}$, $V_{\rm B}$, and $V_{\rm C}$ are the volumes of the
region $A$, $B$, and $C$, respectively, and $\rho^{(j)}$ is the average density
of the $j$th zone.

In Paper I, we also assumed that the metal ejected from a Type Ia supernova
is uniformly dispersed within the SNR radius.
However, Jun, Jones, \& Norman \markcite{jjn1996}(1996) showed that most of
the metal is trapped in the cavity region even in inhomogeneous
medium. 
Thus, in this paper, we assume that the metal
ejected from a Type Ia supernova is trapped in the mixed cavity region. 

After the buoyant mixing, the iron mass in the bored shell 
and mixed cavity region are
\begin{equation}
\label{eq-febs}
m_{{\rm Fe},bs} = m_{bs}\hat{Z}^{(j)} \:,
\end{equation}
\begin{equation}
\label{eq-femc}
m_{{\rm Fe},mc} = m_{\rm Fe}+m_{mc}\hat{Z}^{(j)} \:,
\end{equation}
where $m_{\rm Fe}$ is the iron mass ejected by a supernova and 
$\hat{Z}^{(j)}$ is the average iron abundance of the ISM occupied by the
SNR.

The densities and temperatures determined so far are adopted as
the initial values of the
BSP and MCP for the evolution equations of the phases 
described in \S\ref{sec-beq} as follows : 
\begin{equation}
\label{eq-rhobs}
\rho^{(i,j,bs)}(t_{i,j})=\rho_{bs} \:,
\end{equation}
\begin{equation}
\label{eq-tbs}
T^{(i,j,bs)}(t_{i,j})=T_{bs} \:,
\end{equation}
\begin{equation}
\label{eq-rhomc}
\rho^{(i,j,mc)}(t_{i,j})=\rho_{mc} \:, 
\end{equation}
\begin{equation}
\label{eq-tmc}
T^{(i,j,mc)}(t_{i,j})=T_{mc} \:.
\end{equation}
They are determined uniquely if 
$E_{\rm SN}$, $m_{\rm pro}$, $m_{\rm Fe}$, $k$, $\rho^{(j)}$,
$\hat{T}^{(j)}(=\mu m_{\rm H}P^{(j)}/\rho^{(j)})$, and
$\hat{Z}^{(j)}$ are given, where $m_{\rm pro}$ is the mass of the progenitor 
star and is included by $m_{\rm SNR}$ We fix $E_{\rm SN}$,
$m_{\rm pro}$, and $m_{\rm Fe}$ by giving typical values, that is, 
$m_{\rm pro}=1.4 M_{\sun}$, $m_{\rm Fe}=0.5\rm M_{\sun}$,  
and $E_{\rm SN}=10^{51}$
erg. We also fix $k=0.3$.
The other 
parameters, $\rho^{(j)}$, $\hat{T}^{(j)}$, and $\hat{Z}^{(j)}$ 
are time-dependent and 
determined by solving the evolution
equations of the galaxy described in \S\ref{sec-beq}. 
For examples, for $\rho^{(j)} = 1.67 \times
10^{-26} \rm g \; cm^{-3}$, $\hat{T}^{(j)}
=10^{7}$ K, and $\hat{Z}^{(j)}=1.0 \rm 
Z_{\sun}$, we can estimate each quantity : 
$\rho_{bs}=2.4 \times 10^{-26} \rm g \; cm^{-3}$,
$\rho_{mc}= 8.2 \times 10^{-27} \rm g \; cm^{-3}$,
$T_{bs}=1.5\times10^{7}$ K, 
$T_{mc}=1.5\times10^{7}$ K, $m_{\rm
Fe,bs}/m_{bs}=1\rm Z_{\sun}$, and $m_{\rm Fe,mc}/m_{mc}=2.9 \rm Z_{\sun}$. 
As one can easily calculate from these values, 
the pressures of the MCP, BSP, and
circumference ISM are not identical, that is, they have not been in pressure
equilibrium yet. Therefore, 
we once regard equations (\ref{eq-rhobs}) - (\ref{eq-tmc}) as the values 
at $t=t_{i-1,j}$ and as the initial conditions of the iteration to
derive the values at $t=t_{i,j}$ (see the remark after equation
(\ref{step})). Thus, after the iteration, the MCP, BSP, and
circumference ISM are in pressure
equilibrium. 
Although the cooling time of the MCP is shorter than that of the cavity
phase in Paper I, it is longer than that of the shell phase in Paper
I. This means that more iron ejected by supernova explosion remains
in hot ISM until $t=t_{f}$ than in Paper I. The emission from the MCP
have influence on X-ray spectrum (see \S\ref{sec-anl}). 

The gas and iron 
mass of the BSP and MCP at their birth time 
are given by 
\begin{equation}
M^{(i,j,bs)}(t_{i,j})=\int_{t_{i-1,j}}^{t_{i,j}}L_{\star}^{(j,bs)}(t)dt \: ,
\end{equation}
\begin{equation}
M^{(i,j,mc)}(t_{i,j})=\int_{t_{i-1,j}}^{t_{i,j}}L_{\star}^{(j,mc)}(t)dt \: ,
\end{equation}
\begin{equation}
M_{\rm Fe}^{(i,j,bs)}(t_{i,j})=\int_{t_{i-1,j}}^{t_{i,j}} 
L_{\rm Fe}^{(j,bs)}(t)dt \:,
\end{equation}
\begin{equation}
M_{\rm Fe}^{(i,j,mc)}(t_{i,j})=\int_{t_{i-1,j}}^{t_{i,j}} 
L_{\rm Fe}^{(j,mc)}(t)dt \:,
\end{equation}
where 
$L_{\star}^{(j,bs)}$, $L_{\star}^{(j,mc)}$, $L_{\rm Fe}^{(j,bs)}$, and 
$L_{\rm Fe}^{(j,mc)}$ 
are the mass and iron production rates of the BSP and MCP in the $j$th 
zone,
respectively ; they are given by 
\begin{equation}
\label{star-s}
L_{\star}^{(j,bs)}(t) = r_{\rm SN}(t) m_{bs}(t)M_{\star}^{(j)}(t) \:,
\end{equation}
\begin{equation}
\label{star-c}
L_{\star}^{(j,mc)}(t) = r_{\rm SN}(t) m_{mc}(t)M_{\star}^{(j)}(t) \:,
\end{equation}
\begin{equation}
\label{fe-s}
L_{\rm Fe}^{(j,bs)}(t) = r_{\rm SN}(t) m_{{\rm Fe},bs}(t)M_{\star}^{(j)}(t) \:,
\end{equation}
\begin{equation}
\label{fe-c}
L_{\rm Fe}^{(j,mc)}(t) = r_{\rm SN}(t) m_{{\rm Fe},mc}(t)M_{\star}^{(j)}(t) \:,
\end{equation}
where $r_{\rm SN}(t)$ is the SN Ia rate per unit stellar mass.
The time dependence of the SN Ia rate is
assumed to be $r_{\rm SN}(t) \propto t^{-0.5}$ (David, Forman, \& Jones 
\markcite{dfj1990} 1990).
The normalization is given by the SN Ia rate at $t=t_{f}$,
which is specified later. 

\subsubsection{Can Magnetic Field Suppress Mixing of the Phases?}
\label{sec-mag}
\indent

Nulsen \markcite{n1986}(1986) pointed out that 
perturbations can be pinned in cooling
flows when magnetic stresses are able to suppress their relative
motions. This occurs when the Alf{\'v}en speed, $v_{\rm A}$ is larger
than the terminal velocity, $v_{t}$ 
defined by the force
balance between ram pressure and the excess gravity.
Hattori et al. \markcite{hyh1995}(1995) confirm this numerically; 
they show that magnetic field can
suppress relative motion in a non-linear perturbation and can help the
growth of thermal instability as long as the size of perturbation is 
smaller than the critical value
\begin{equation}
\label{eq-crit}
\lambda_{\rm crit} = 1.5\left(\frac{\delta}{10}\right)^{-1}
		\left(\frac{\beta}{350}\right)^{-1}
		\left(\frac{L}{5 \;\rm kpc}\right)\rm pc \:,
\end{equation}
where $\beta$ is the ratio of the gas pressure to the magnetic
pressure, and 
$L$ is the scale height of
pressure.
The density contrast, $\delta$, is defined as 
\begin{equation}
\delta = \frac{\rho-\rho_{b}}{\rho_{b}} \:,
\end{equation}
where $\rho$ and $\rho_{b}$ are the density of the perturbation and
unperturbed flow, respectively.

For the case of a temperature of $10^{7}$ K and a density of
$\rho_{b}= 1.67\times 10^{-26} \: \rm g \:cm^{-3}$, 
and there is a $1\mu G$ 
magnetic field, one finds that $\beta=350$ and $r_s=70$ pc ;  
in the following argument in this subsection, we will use these
parameters and take $L=5$ kpc. 
When $k=0.3$, spatial scale lengths of cavity and shell region respectively 
are 50 pc and
$20/2=10$ pc, following the definition of Hattori et al. 
\markcite{hyh1995}(1995). 
Since the initial value of $\delta$ for the shell region is 
less than one, the critical wavelength is 
$\lambda_{\rm crit} > 15$ pc. This indicates that the
shell region can be supported by magnetic field before the cooling
becomes efficient. On the contrary, one can show that the 
cavity region, before mixing, cannot be supported, 
if equation (\ref{eq-crit}) can be
applied to less dense perturbation for $|\delta| \sim 1$ ; 
it will move upwards and mix with
the ambient medium as we assumed in \S\ref{sec-sn}.

As each phase cools, the density contrast increases. 
Hattori \& Habe
\markcite{hh1990}(1990) show that if magnetic tension supports a
perturbation until $\delta > 10$, the gas in the perturbation can cool 
to $10^{4}$ K. From equation (\ref{eq-crit}), one finds $\lambda_{\rm
crit} = 1.5$ pc when $\delta = 10$. 
Since the spatial scale length of gas blobs composing a mass-loss phase 
is typically $\sim 1$ pc (Mathews \markcite{m1990}1990), 
the blobs can cool to $10^{4}$ K without mixing and drop out of 
hot ISM. On the other hand, since the spatial scale lengths of mixed
cavity region and bored shell region are 50 and 10 pc, respectively,
they cannot cool to $10^{4}$ K without mixing 
unless they fragment into small pieces
before the cooling becomes effective ($\gtrsim 10^{7-8}$ yr).
However, Jun et al. \markcite{jjn1996}
(1996) show that a SNR in 
inhomogeneous medium has complicated structure ; thus, we do not think that
each region 
evolves as one gas blob for a long time. Moreover, the estimation
of Hattori \& Habe (1990) does not include the effect of metal
abundance fluctuation which makes gas blobs with higher metal
abundance cool faster, and makes the value of $\delta$ until which 
magnetic tension should support them smaller. 
Furthermore, since a
SNR amplifies magnetic field (Jun \& Norman
\markcite{jj1996}1996), 
$\lambda_{\rm crit}$ may
be large around it. Thus, we assume in \S\ref{sec-anl}
that the MCP and BSP also comove with
ambient ISM, although we consider an additional 
model in which the MCP and BSP mix
with ambient ISM in wider region.  

The above estimations may be too simple ; strength variation and structure of
magnetic fields may affect the evolutions of gas blobs. However, since 
a full treatment of convection in inhomogeneous medium would be very 
difficult, we consider
comoving flows as a first-step of the research.

\subsection{EVOLUTION OF THE PHASES AND THE GALAXY}
\label{sec-beq}
\indent

The energy equation for the $(i,j,\alpha)$-th phase for $t > t_{i,j}$ is  
given by 
\begin{equation}
\label{energy}
\frac{\rho^{(i,j,\alpha)}(t)}{\gamma-1}\frac{d}{dt}\left(\frac{k_{\rm B}
T^{(i,j,\alpha)}(t)}{\mu
m_{\rm H}}\right)
-\frac{k_{\rm B}T^{(i,j,\alpha)}(t)}
{\mu m_{\rm H}}\frac{d}{dt}\rho^{(i,j,\alpha)}(t)
=-(n_{e}^{(i,j,\alpha)})^{2}
\Lambda(Z^{(i,j,\alpha)},T^{(i,j,\alpha)}(t)) \:,
\end{equation}
where 
$n_{e}^{(i,j,\alpha)}$ is the electron density 
of the $(i,j,\alpha)$-th phase and $\Lambda$ 
is the cooling function approximated by
\begin{eqnarray}
\label{eq-cool}
\Lambda(Z^{(i,j,\alpha)}, T^{(i,j,\alpha)}) 
 & = &\left[2.1 \times 10^{-27}
      \left(1 + 0.1\frac{Z^{(i,j,\alpha)}}{\rm Z_{\sun}}\right) 
      \left(\frac{T^{(i,j,\alpha)}}{\rm K} \right)^{0.5}\right.
      \nonumber \\
 & + &\left.8.0 \times 10^{-17}
      \left(0.04 + \frac{Z^{(i,j,\alpha)}}{\rm Z_{\sun}}\right)
      \left(\frac{T^{(i,j,\alpha)}}{\rm K}\right)^{-1.0}\right] 
      \nonumber \\
 &   & (\rm ergs \: cm^{-3} s^{-1}) \:.
\end{eqnarray}
This is an empirical formula derived by fitting to the cooling curves
calculated by B\"{o}hringer \& Hensler
(1989). 

In our model, we assume that all existing phases in a certain zone 
have same pressure.
Thus, for $t_{i,j}<t_{l,j}$, 
\begin{equation}
\label{peq}
\rho^{(i,j,\alpha)}(t_{l,j})\frac{k_{\rm B}T^{(i,j,\alpha)}(t_{l,j})}
{\mu m_{\rm H}}
 = P^{(j)}(t_{l,j}) \:.
\end{equation}

We assume that the ISM is in hydrostatic equilibrium, that is,
\begin{equation}
\label{rho}
\frac{\Delta P^{(j)}}{\Delta R_{j}}
=-\rho^{(j)}\frac{G M(R_{j})}{(R_{j})^{2}} \:,
\end{equation}
\begin{equation}
\rho^{(j)}(t_{l,j})=\frac{M_{\rm g}^{(j)}(t_{l,j})}{V^{(j)}(t_{l,j})} \:,
\end{equation}
where $M_{\rm g}^{(j)}$ and $V^{(j)}$ are respectively the total gas 
mass and volume of the $j$th zone, $\Delta P^{(j)}=P^{(j+1)}-P^{(j)}$, 
and $\Delta R_{j}=R_{j}-R_{j-1}$. We define $P^{(j)}$ as the pressure
just outside $R_{j-1}$.

The gravitational mass within $R$ is given by
\begin{equation}
M(R)=\int_{0}^{R}4\pi R^{2}(\rho_{\star}+\rho_{h})dR \:.
\end{equation}

The total gas mass, the iron mass, and the gas volume of the $j$ the
zone are the 
summation of those of each phase ; 
\begin{equation}
\label{mg}
M_{\rm g}^{(j)}(t)=\sum_{\alpha}
             \sum_{i,\: T^{(i,j,\alpha)}>T_{\rm crit}}M^{(i,j,\alpha)}(t) ,
\end{equation}
\begin{equation}
\label{mfe}
M_{\rm g,Fe}^{(j)}(t)=\sum_{\alpha}
          \sum_{i,\: T^{(i,j,\alpha)}>T_{\rm crit}}
M_{\rm Fe}^{(i,j,\alpha)}(t) ,
\end{equation}
\begin{equation}
\label{volume}
V^{(j)}(t)=\sum_{\alpha}\sum_{i,\: T^{(i,j,\alpha)}>T_{\rm crit}}
         V^{(i,j,\alpha)}(t) ,
\end{equation}
where 
\begin{equation}
V^{(i,j,\alpha)}(t) = 
M^{(i,j,\alpha)}(t)/\rho^{(i,j,\alpha)}(t) \:. 
\end{equation}
As mentioned above, the phases whose
temperatures are below $T_{\rm crit} (= 10^{5}\rm K)$ are not included 
in the summation.

Note that part of the SNRs is composed of pre-existing phases ; 
the masses of the pre-existing phases
are reduced by the occupation by the SNRs.
Thus, the mass and iron mass 
of the pre-existing $(i,j,\alpha)$-th phase at $t=t_{l,j}$ 
$(t_{i,j}<t_{l,j})$ are given by 
\begin{equation}
\label{red}
M^{(i,j,\alpha)}(t_{l,j}) = M^{(i,j,\alpha)}(t_{l-1,j})-
\left. M_{{\rm SNR},l}^{(j)}
\frac{V^{(i,j,\alpha)}(t)}{V^{(j)}(t)} 
\right|_{t=t_{l-1,j}} \:,
\end{equation}
\begin{equation}
\label{redfe}
M_{\rm Fe}^{(i,j,\alpha)}(t_{l,j}) = M_{\rm Fe}^{(i,j,\alpha)}(t_{l-1,j})-
\left. M_{{\rm SNR},l}^{(j)}
\frac{M_{\rm Fe}^{(i,j,\alpha)}(t)}{M^{(i,j,\alpha)}(t)}
\frac{V^{(i,j,\alpha)}(t)}{V^{(j)}(t)} 
\right|_{t=t_{l-1,j}} \:,
\end{equation}
respectively, where $M_{{\rm SNR},l}^{(j)}$ is the mass occupied 
by SNRs during
$t_{l-1,j}<t<t_{l,j}$, and is given by
\begin{equation}
M_{{\rm SNR},l}^{(j)}=\int_{t_{l-1,j}}^{t_{l,j}}
\rho^{(j)}(t) V_{S}
r_{\rm SN}(t)M_{\star}^{(j)}(t) dt \:,
\end{equation}
where $V_{\rm S}=V_{\rm A}+V_{\rm B}+V_{\rm C}+V_{\rm D}$.

We take the sound crossing time as the time-steps for the calculations, 
\begin{equation}
\label{step}
t_{l+1,j}-t_{l,j}= \Delta R_{j}(t_{l,j})
             \left(\gamma \frac{k_{\rm B}\hat{T}^{(j)}}
             {\mu m_{\rm H}}\right)^{-1/2}
\end{equation}

We derive $\rho^{(i,j,\alpha)}(t_{l,j})$ and $T^{(i,j,\alpha)}(t_{l,j})$ from
$\rho^{(i,j,\alpha)}(t_{l-1,j})$ and 
$T^{(i,j,\alpha)}(t_{l-1,j})$ by iterating 
Eqs.(\ref{energy}),
(\ref{peq}), (\ref{rho}), (\ref{mg}), (\ref{volume}), and 
(\ref{red})
until they converge.

\section{MODEL RESULTS}
\label{sec-anl}
\indent

In this section, we present results of the evolution equations
described above. 
We reduce the number of the free parameters by
giving typical fixed values to some of them.
First, for the parameters regarding the whole galaxy, 
we take $t_{0}=0.5$ Gyr, $t_{f}=10$ Gyr. 

We assume that the gas distribution and abundance distribution 
at $t=t_{0}$ are 
\begin{equation}
\rho_{gas}(R) =  \rho_{0 gas}[1+(R/R_{a\star})^{2}]^{-3/2} \:, 
\end{equation}
\begin{equation}
\label{eq-zml0}
Z_{\rm 0}(R)=Z_{\rm 0}(R_{e})(R/R_{e})^{-1/2} \:
\end{equation}
for $R<R_{t}$, respectively. 
The normalization, $\rho_{0 gas}$, is determined by assuming 
that the total gas mass at $t=t_{0}$ is 1/30 of the total stellar mass.
The normalization, $Z_{\rm 0}(R_{e})$, is a parameter. 
We give the initial zone radius so that enough spatial resolution is
obtained at $t=t_{f}$. The initial number of the zone, $m$, is 30.
We can write 
\begin{equation}
V^{(0,j,0)}(t_{0})=V^{(j)}(t_{0})=\frac{4\pi}{3}[R_{j}(t_{0})^{3}
-R_{j-1}(t_{0})^{3}] \:, 
\end{equation}
\begin{equation}
M^{(0,j,0)}(t_{0})=M_{\rm g}^{(j)}(t_{0})
                =\int_{R_{j-1}(t_{0})}^{R_{j}(t_{0})}
		4 \pi R^{2} \rho_{gas}(R)dR \:,
\end{equation}
\begin{equation}
M_{\rm Fe}^{(0,j,0)}(t_{0})=M_{\rm Fe,g}^{(j)}(t_{0}) \:.
\end{equation}
We take the temperatures of the zero-phases 
\begin{equation}
T^{(0,j,0)}(t_{0})=T_{\star}(R_{j}(t_{0})) \:.
\end{equation}

We solve the basic 
equations described 
in \S \ref{sec-2} for the galaxy models whose details
are given in Table 1. 
The SN Ia rate is normalized by the values at $t=t_{f}(=10\rm Gyr)$ 
shown in Table 1. The values are expressed 
in units 
of SNu, that is, the number of 
SNe per $10^{10} h^{-2} \rm L_{B\sun}$ per 100 yr ($H_{0}=100h \rm km 
\: s^{-1} \: Mpc^{-1}$ ; we set $h=0.5$). The ratio of the mass to
luminosity is taken to be
$8M_{\sun}/L_{\sun}$, and the rate for SN Ia is normalized by using
this ratio.
Models A3-A6 and B3-B6 are calculated to see the effect of 
metal-abundance distribution for the mass-loss gas.
In these models, the mass-loss phase born in
each time-step is divided into two phases with equal masses and 
different abundances corresponding to the two figures 
for $Z_{\rm ML}(R_{e})$ (Table 1).
On the other hand, in models A1 and A2, 
each component has the same
abundance or the mass-loss gas has one kind of the abundance. 
In model A5, shell and cavity regions mix into circumference ISM of 26
times their volumes, respectively. 
This means that shell and cavity region 
mix into circumference ISM in 
wider region than other models. 

Figure 2(a) shows the time evolution of 
$R_{j}$ ($1\leq j \leq m$) in
models A1 and A2 ; 
this figure shows the influence of SN Ia on the gas 
evolution. Cooling flow is 
established in the inner
region. Note that in other models, 
the cooling flow is also established. 
Although the supernova rate of model A2 is larger than that 
of model A1, the inward velocity in model A2 is not reduced. 
This fact indicates that supernovae are {\em not} effective heating 
sources of the ISM in the {\em inner} region of the galaxy.
The reason is that the BCP and MCP cool 
effectively owing to their high metal abundance
(see \S\ref{sec-sn}), and that they have large thermal energies 
when they are born ; the sum of their energy is 
\begin{equation}
E_{bs}+E_{mc} =   
          E_{\rm SN}+\frac{3 k\hat{T}}{2\mu m_{\rm H}}
          \rho V_{\rm S} > E_{\rm SN}
\: 
\end{equation}
except for models A5 and B5.
This indicates that more energy than that supplied by 
supernova explosion
radiates from the BSP and MCP in a short time. Thus, supernovae 
cannot heat the ISM effectively. 
In {\em outer} region, the outflow velocity in model A2 is 
larger that in
model A1 ; supernovae {\em are} 
effective as
heating sources.  
In this region, cooling times of the BSP and MCP are so long that 
they do not radiate their thermal energy so much. 
Thus, the energy released by supernovae is transfered to the
circumferential ISM before the cooling. 

In Figure 2(b), outward velocity of model A5 is larger than that of
model A3. This is because mixing of SNRs makes their cooling time
longer. Therefore, the energy transition between SNRs and circumferential ISM 
is more effective in model A5. 

Figures 3 - 5 show the distributions of the density, 
luminosity averaged 
temperature, and metal abundance at $t=t_{f}$ (at $t=9.4$ Gyr for model
A5 because the radius of the innermost zone is $\sim 20$ kpc at $t=t_{f}$). 
We confirmed that, for models in Paper I, the luminosity averaged 
temperatures are almost same
as those derived by spectral simulations which we did in Paper I. 
On the contrary, the luminosity averaged metal abundances are not
consistent with those derived by the spectral simulations.
Thus in this section, we simply use $Z_{m}=M_{\rm
g,Fe}^{(j)}/M_{\rm g}^{(j)}$ as the metal abundance, and later,
derive X-ray spectrum of the model galaxies. 
The units of metal abundance is 
the solar abundance $\rm Z_{\sun}$, $1.7 \times 10^{-3}$.
Since our model does not have enough spatial resolution in the
central region of the galaxy, we infer the distributions of the 
density, luminosity averaged temperature, and metal abundance by
presenting the values of the zones which disappear just before
$t=t_{f}$. The values 
shown in Table 2 are the ones when the zone
radii are $\sim 0.3$ kpc.
For model A2, we stopped calculating the evolution of the innermost zone  
and deleted it before its radius decreases to 0.3 
kpc. This is because  
masses of 
some phases in the innermost zone become
negative by equation (\ref{red}), because of 
large $M_{{\rm SNR},l}^{(j)}$ and 
density contrast among phases. In order to overcome 
this defect of our 
model, the exact treatment of the evolution of supernova 
remnants in
inhomogeneous medium is required. However, we think this deletion 
does not affect the evolution of the outer zones, because when we
stop the calculation, the volume of the innermost zone is far 
smaller 
than those of the outer zones. 

Figure 3 show that 
the density varies crudely as $\rho \propto R^{-1.5}$ for $R \gtrsim 
1$ kpc.
The distributions of
density are almost independent of the models. 
This reflects that the density 
distributions are mainly
decided by energy injection rate and vary as $\propto
\rho_{\star}^{1/2}$ (Sarazin, \& White \markcite{sw1987}1987). 
The high central density of models A5 and B5 
reflects small density contrast
among the phases in the innermost zones (Table 2).

The temperature gradually rises toward the galactic center 
(Figure 4). This
temperature rise is caused by adiabatic heating owing to 
the gas inflow
(Figure 2) ; in our models, the phases with low-cooling rates are 
brought to the center and compressed,  
because high-cooling phases drop out and 
the volume of the inner zones reduces. In the central region,
the 
temperature falls down (Table 2). 
In this region, all phases start to
cool because of high pressure. 

The metal abundances, $Z_{m}$, are very different among the models
A1-A3 
(Figure 5(a)). 
We show the metal abundance distribution of mass-loss gas 
when $Z_{\rm ML}(R_{e})=1 Z_{\sun}$ 
(equation (\ref{eq-zml})) in Figure 5 for reference. 
In model A2, the high supernova rate makes the metal
abundance high. This high metal abundance is inconsistent with the
observations (Awaki et al. \markcite{amt1994}1994 ;  Loewenstein et al.
\markcite{lmt1994}1994 ; Matsushita 
et al. \markcite{mma1994}1994 ; Mushotzky et al. \markcite
{mla1994}1994 ; 
Kim \& Fabbiano \markcite{kf1995}1995 ; Davis \& White
\markcite{dw1996}1996 ; Matsumoto et
al. \markcite{m1996}1996). This may indicate
that the real supernova rate is 
lower than that derived by the observation ($\sim 0.2$ SNu ; 
Tammann \markcite{t1982}
1982). 
In the central region, high-abundance phases
generally cool and drop out of hot ISM faster 
than other phases, although it depends on 
their temperatures and densities. 
This makes 
the average metal abundance in that region small.
However, since the negative metal abundance gradient of the stars 
or mass-loss gas cancels this effect, the metal abundance of the 
hot ISM is constant or rises toward 
the center of the model galaxy except for model A6 and B6 
(Figure 5). 
Since this result
depends on the assumption of equation (\ref{eq-zml}), the observation
of the metal abundance of the stars beyond the effective radius is
encouraged. 
Moreover, gas flows inward from the 
outer region where the
metal abundance of mass-loss gas is small. 
This means that as gas ejection rate of the stellar system 
decreases, mass fraction of low-abundance phases born in the
outer region increases in a galaxy.
In the outer
region, the selective cooling is not effective and the 
metal abundances of the hot ISM directly reflects the metal
injection from stars including supernovae.   

The metal abundance in models A3 is smaller than
those in models A1 (Figure 5(a)), 
because the mass-loss phases with higher metal 
abundance cool faster than those with lower metal abundance. 
Compared with model A3, model A4 shows higher abundance (Figure 5(a) 
and 5(b)). 
In model A4, part of high abundance gas of the zero-phase is occupied by  
SNRs which survive until $t=t_{f}$. 
In model A6 and B6 , mass-loss phases with low metal abundance dominate in the 
central region at $t=t_{f}$.

Although the metal abundances of the models except for A2 
is low, they cannot be compared directly 
with the {\em ASCA} results ($\lesssim 0.5\rm Z_{\sun}$). 
Since the density of phases with high abundance are generally high,
emission from them can affect X-ray spectrum. 
Therefore, we simulate the spectra at $t=t_{f}$, as we did in Paper I, 
for models A3
- A6. We simulate the spectra of innermost zones and those of zones 
at $\sim 20$ kpc. Although we do not consider projection effect,
contamination of outer zones is less than 20\% for the innermost
zones. For $\gtrsim 20$ kpc, the variation among zones are little
except for densities. The distance to the model galaxies is 
assumed to be 15
Mpc. We use XSPEC package (version 9.00) and response function of {\em 
ASCA} Solid-State Imaging Spectrometer (SIS). The model spectrum are fitted 
by Meka model (Mewe, Gronenschild, and van den Oord\markcite{mgv1985} 
1985 ; Mewe, Lemen, and van den
Oord\markcite{mlv1986} 1986 ; Kaastra\markcite{k1992} 
1992). The method is described in Paper I. 

The results are shown in Figure 6. 
Because of the low luminosity, $Z_{sp}$ at $\sim 20$ kpc has an error
of $\sim 0.4 Z_{\sun}$.
We present the spectrum of the
innermost zone of model A4 as an example (Figure 7). 
The central metal abundance
derived by spectral fitting, $Z_{sp}$ is larger than $Z_{m}$ 
for models A3 and A4. This is because that 
MCPs affect the X-ray spectrum by 
their strong metal emission. The number of MCP
which can survive until $t=t_{f}$ is small compared with those of 
other phases;
for example, for the innermost zone in 
model A4 the ratio is $MLL : MLH : BSP : MCP = 161 :
14 : 146 : 97$, where $MLL$ and $MLH$ refer to mass-loss phase with
low and high metal abundance, respectively. On the contrary, the
luminosity ratio is $MLL : MLH : BSP : MCP 
\sim 2 : 1 : 4 : 4$. 
Note that for the zone at $\sim 20$ kpc, the luminosity ratio is 
$MLL : MLH : BSP : MCP 
\sim 2 : 3 : 2 : 2$. In model A5, $Z_{sp}\sim Z_{m}$ in the galactic 
center, because of small 
abundance fluctuation among the phases. However, the luminosity of MCP
and BSP dominates that of mass-loss phase as models A3 and A4.  

In the central region, the derived metal abundances, $Z_{sp}$, are
larger than the {\em ASCA} results (Mushotzky et
al. \markcite{m1994}1994) except for 
model A6, although $Z_{sp} \lesssim Z_{\rm ML}$ (Figure 6). 
This may indicate that our model is too simple to predict
metal abundance quantitatively. Alternatively, supernova explosion rate
in elliptical galaxies may be $<0.01$ SNu ; the central abundance
deficit of NGC 4472 (Irwin \& Sarazin \markcite{ir1996}1996) 
may correspond to that in models A6 and B6.
As discussed above, our model predicts that in
a central region of a galaxy, 
iron line emission from SNRs of Type Ia supernovae, if any, 
should be prominent. That is, observed 
[Si/Fe] and [Mg/Fe] should decrease towards the 
center. 

The X-ray luminosities of the model galaxies are $\sim 10^{41-42}
\rm erg \: s^{-1}$ for $t \sim 10^{10}$ yr ; the luminosities 
fluctuate because of the small number of the zone.

\section{SUMMARY AND CONCLUSIONS}
\indent

In this paper, we have presented a model of the 
evolution of the hot gas for
elliptical galaxies after galactic wind period 
under the assumption that the gases ejected from
stars do not mix with the circumferential ISM globally.
The ejected gases 
evolve separately according to their birth time, position, and 
origin. We
considered three origins of the ejected gas, that is, shell and cavity
of supernova remnants and mass-loss gas. Furthermore, we considered the
floating of the cavity and subsequent mixing by Rayleigh-Taylor 
instability. 

The main results and conclusions can be summarized as follows:

\begin{enumerate}
\item 
The model predicts that the supernovae are {\em not} effective as heating
sources of the ISM in the {\em inner} region of galaxies 
after the galactic wind stops. 
In the inner region, supernova remnant can cool 
rapidly because of
their high density and/or metal abundance. 
Since the remnants initially have large thermal energy, 
the energy ejected by supernova
explosions is radiated and supernovae do not heat up the ISM. 
Thus, cooling
flow is established even if supernova rate is large. 
In the {\em outer} region of the galaxies, the cooling time 
of the remnants is long. Thus, most of energy 
ejected by supernova
explosions is not radiated and it is transfered into the circumferential 
ISM. Mixing of SNRs with ambient ISM makes this transfer more
effective. 

\item 
In a inner region of a galaxy, 
the present iron abundance of the hot ISM can be less
than that of the mass-loss gas or 
stars if the supernova rate is small, because 
the phases with higher metal 
abundance generally cool faster and gas inflows from 
outer region where the
metal abundance of the mass-loss gas is small.
However, the spectral simulations show that predicted metal abundances
are still larger than the ones observed
by {\em ASCA} in the central region, 
if the present supernova rate is $\gtrsim 0.01$ SNu.
In the outer region where the selective cooling is ineffective, 
metal abundance of the ISM directly 
reflects that of the gas ejected from stars. 
Our model predicts that iron line emission by SNRs is prominent in the 
central region, and that
[Si/Fe] and [Mg/Fe] decrease towards the galactic 
center. 
\end{enumerate}

\acknowledgments

We thank H. Kodama, N. Gouda, H. Sato, 
M. Hattori, Y, Ikebe, H. Susa, and T. Kan-ya 
for useful discussion and 
comments. We appreciate K. Koyama for using a computer.  
We are also grateful to H. Matsumoto and E. Rokutanda for providing
{\em ASCA} data. This work was supported in part by the JSPS Research
Fellowship for Young Scientists.

\vspace{10mm}

\noindent
{\em Note added in proof}

\vspace{1mm}

A recent {\em ASCA} result shows that the metal abundances in central regions
of elliptical galaxies are $\sim 1\; \rm Z_{\sun}$ 
(\markcite{m1997}Matsushita 1997).

\newpage

\thispagestyle{empty}

\section*{Figure Captions}

\figcaption[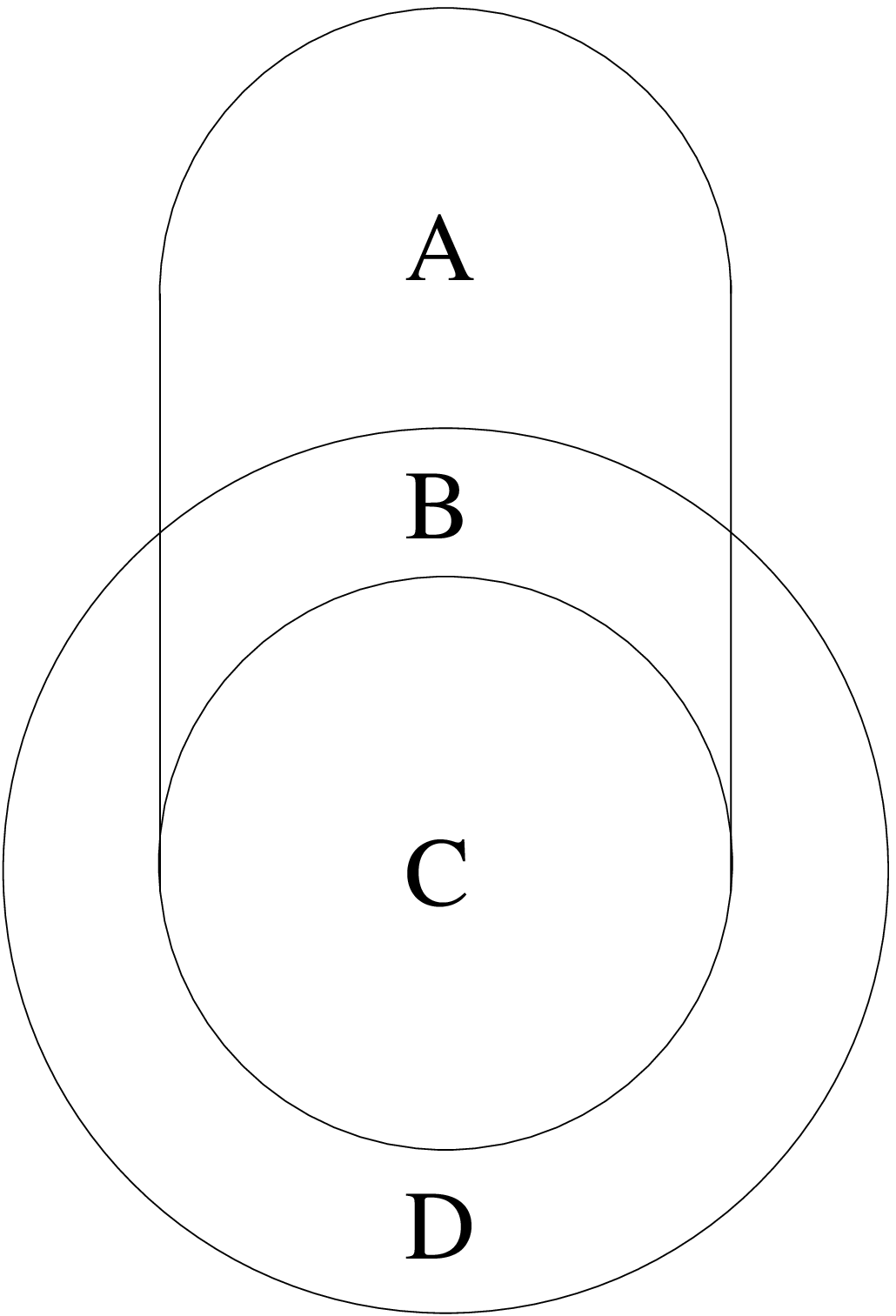]{The schematic figure of a supernova
remnant.}

\figcaption[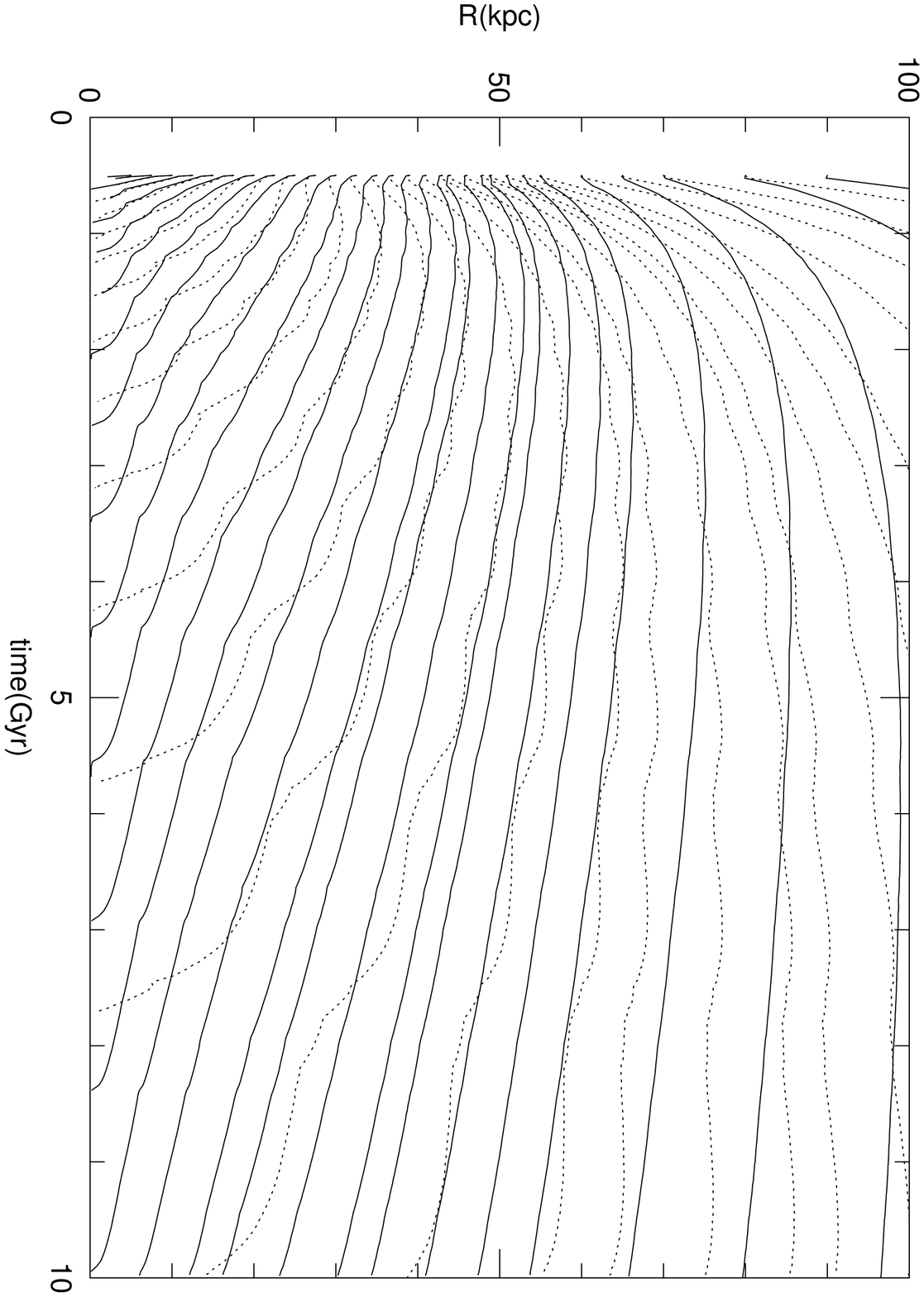]{The evolutions of zones for
(a) model A1 (solid line) and A2 (dotted line) 
(b) model A3 (solid line) and A5 (dotted line).}

\figcaption[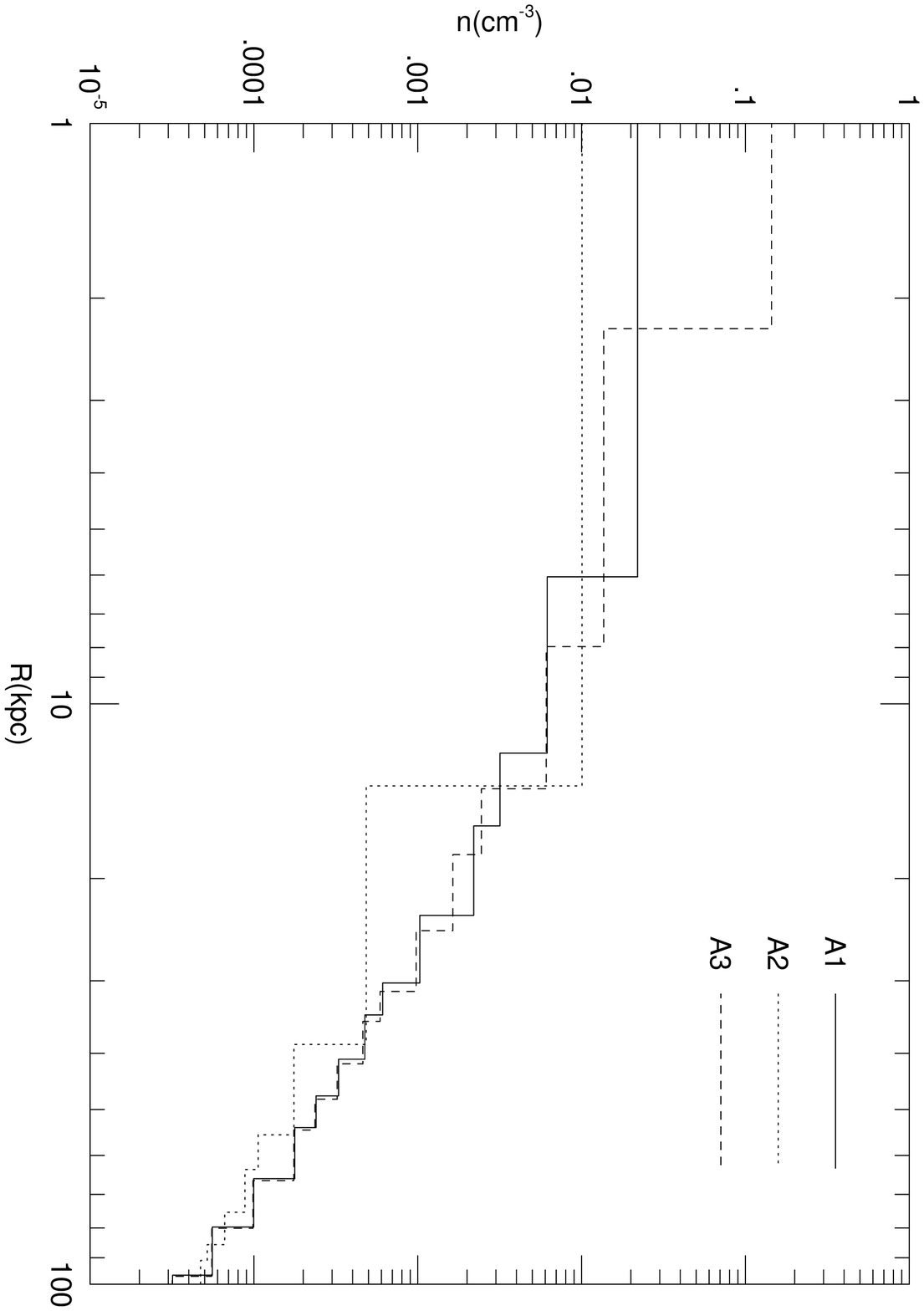]{The present distribution of 
density. 
(a) A1 - A3, (b) A4 - A6, and (c) B4 - B6.}

\figcaption[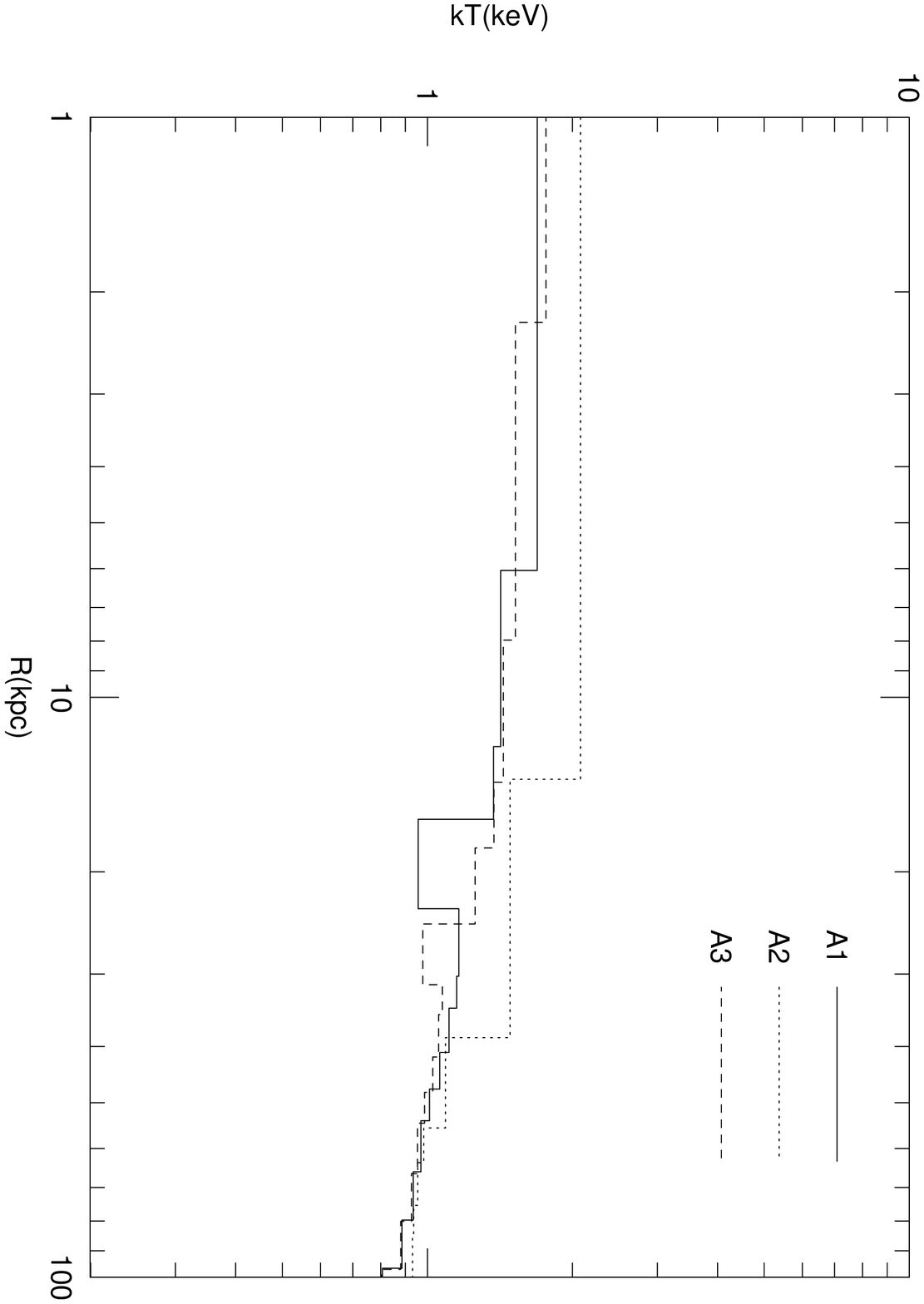]{The present distribution of 
temperature. 
(a) A1 - A3, (b) A4 - A6, and (c) B4 - B6.}

\figcaption[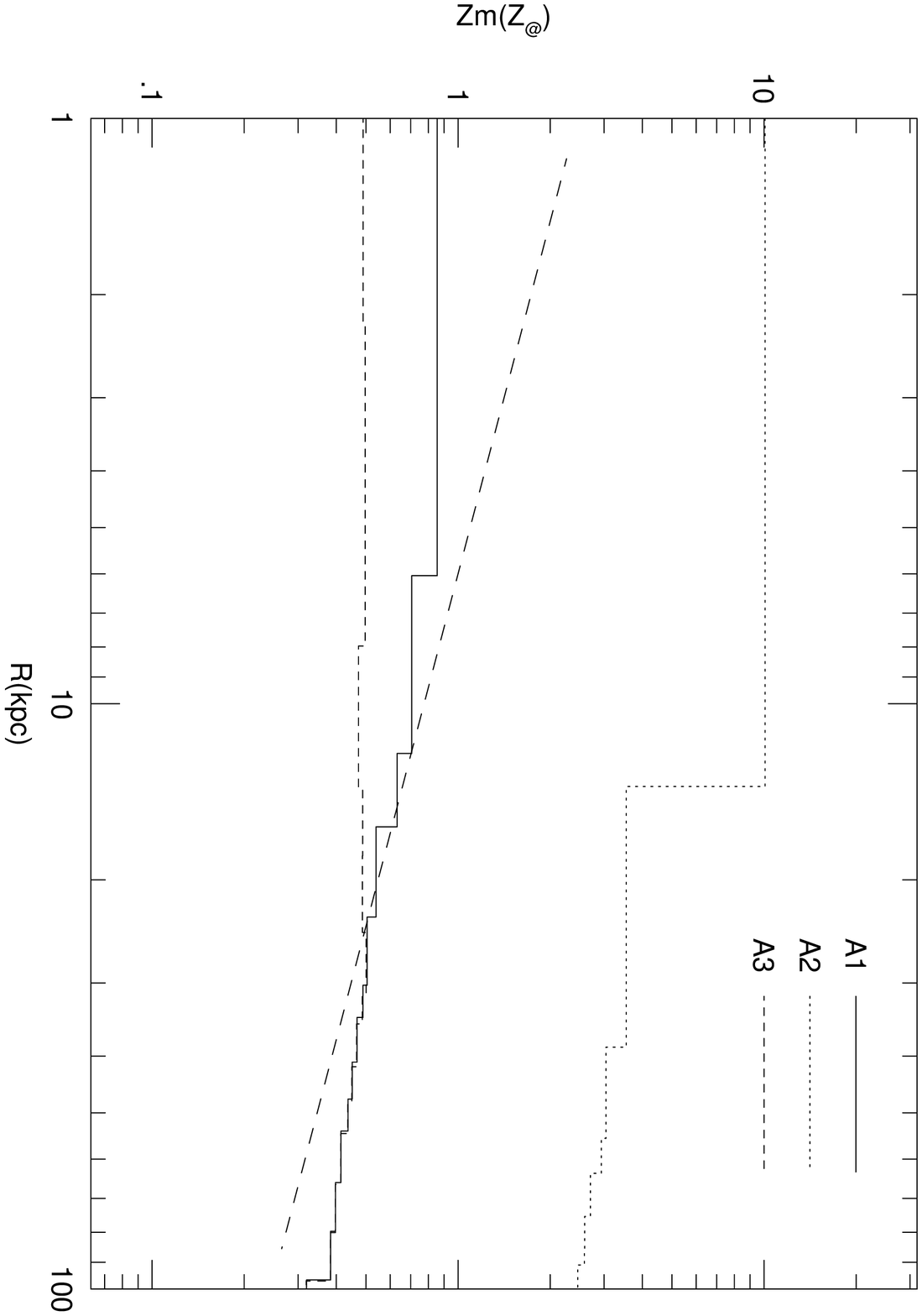]{The present distribution of 
abundance ($Z_{m}$). 
(a) A1 - A3, (b) A4 - A6, and (c) B4 - B6.
The long-dashed line shows the metal 
abundance of the mass-loss gas when $Z_{\rm ML}(R_{e})=1.0$.}

\figcaption[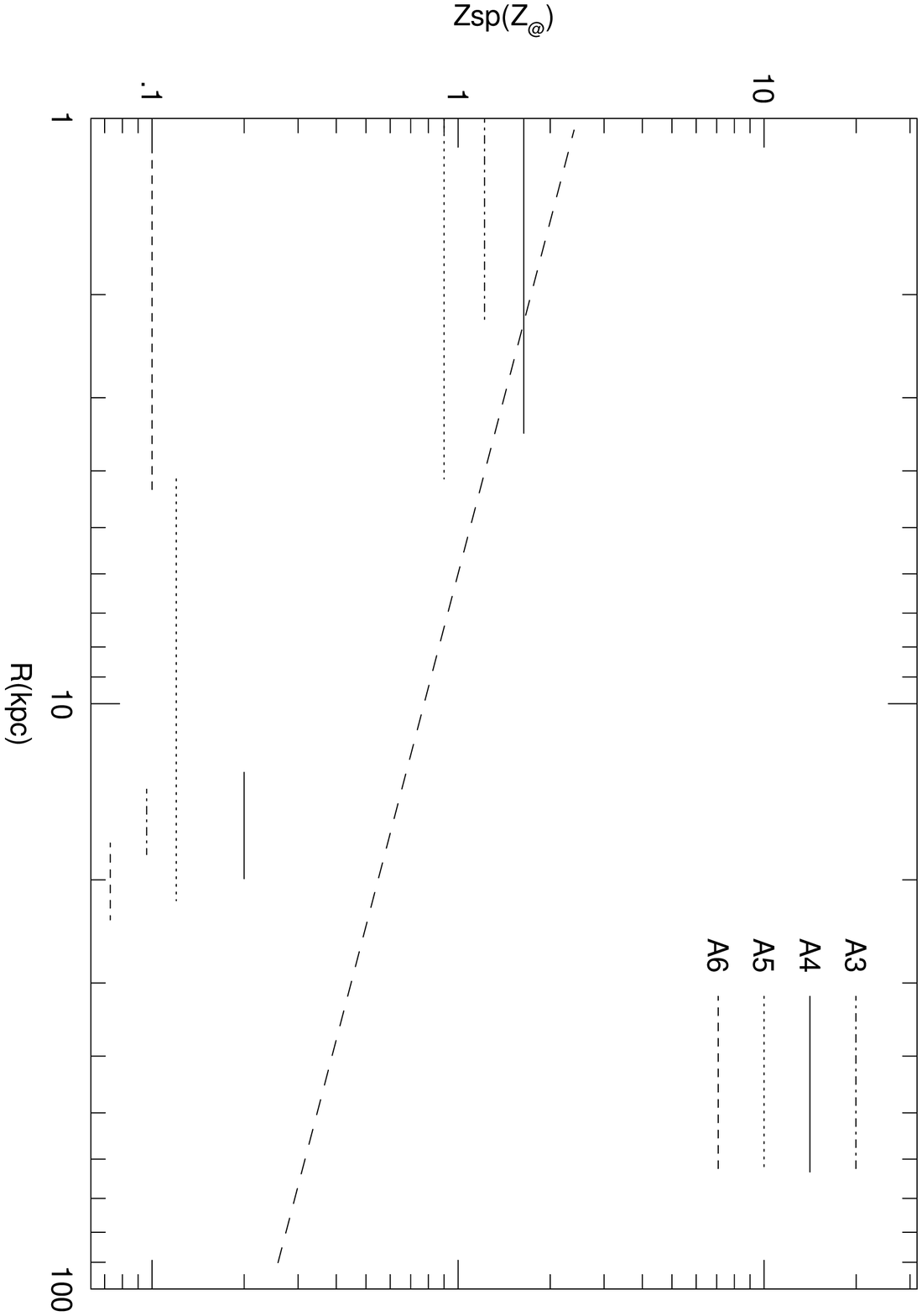]{The present distribution of abundance ($Z_{sp}$)
for models A3 - A6. The long-dashed line shows the metal 
abundance of the mass-loss gas when $Z_{\rm ML}(R_{e})=1.0$.}

\figcaption[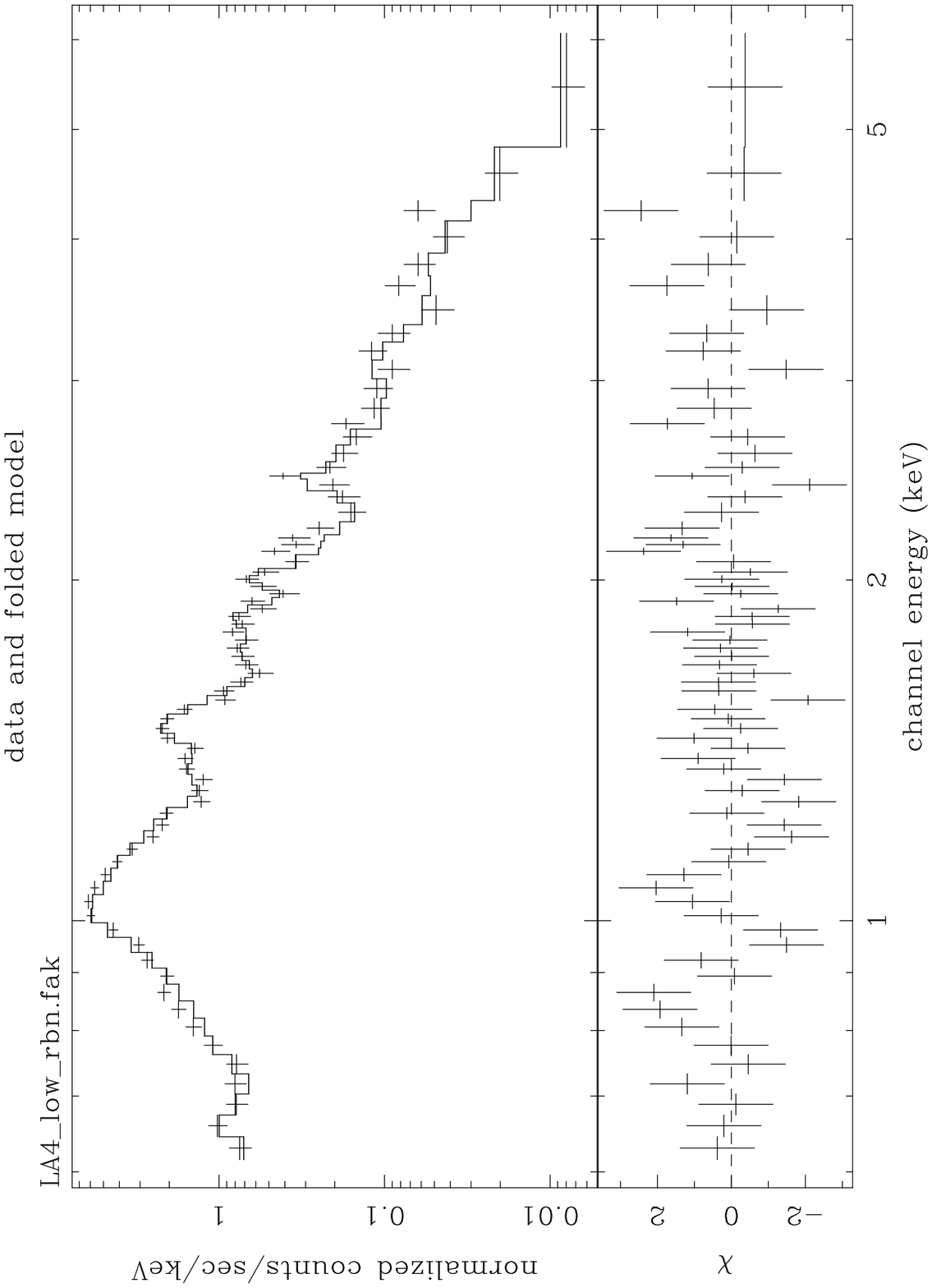]{Simulated X-ray spectra observed 
with the {\em ASCA} SIS for model A4. 
The line shows the best
fitting (two Meka plasma + absorption column density).
}

\newpage



\makeatletter
\def\jnl@aj{AJ}
\ifx\revtex@jnl\jnl@aj\let\tablebreak=\nl\fi
\makeatother


\begin{deluxetable}{cccccccccc}
\tablewidth{0pc}
\tablecaption{Properties of the Model Galaxies}
\tablehead{
\colhead{Model}           & \colhead{$R_{a \star}$}       &
\colhead{$\rho_{0 \star}$}& \colhead{$R_{a h}$}           &
\colhead{$\rho_{0 h}$}    & \colhead{$R_{e}$}             &
\colhead{$R_{t}$}	  & \colhead{SN rate}             &
\colhead{$Z_{\rm ML}(R_{e})$}&\colhead{$Z_{0}(R_{e})$}
\\
\colhead{}                & \colhead{(kpc)}               &
\colhead{($10^{-21}\rm g\;cm^{-3}$)}& \colhead{(kpc)}     &
\colhead{($10^{-23}\rm g\;cm^{-3}$)}& \colhead{(kpc)}     &     
\colhead{(kpc)}           & \colhead{(SNu)}		  &
\colhead{($Z_{\sun}$)}    & \colhead{($Z_{\sun}$)}
}

\startdata
A1   & 0.50 & 7.20 & 5.0 & 1.75 & 6.0 & 100.0 & 0.01 & 1.0   &1.0\nl
A2   & 0.50 & 7.20 & 5.0 & 1.75 & 6.0 & 100.0 & 0.2  & 1.0   &1.0\nl
A3/A5& 0.50 & 7.20 & 5.0 & 1.75 & 6.0 & 100.0 & 0.01 &0.2,1.8&1.0\nl
A4   & 0.50 & 7.20 & 5.0 & 1.75 & 6.0 & 100.0 & 0.01 &0.2,1.8&4.0\nl
A6   & 0.50 & 7.20 & 5.0 & 1.75 & 6.0 & 100.0 & 0    &0.2,1.8&1.0\nl
B4   & 0.25 & 21.8 & 2.5 & 5.30 & 3.0 & 50.0  & 0.01 &0.2,1.8&4.0\nl
B5   & 0.25 & 21.8 & 2.5 & 5.30 & 3.0 & 50.0  & 0.01 &0.2,1.8&1.0\nl
B6   & 0.25 & 21.8 & 2.5 & 5.30 & 3.0 & 50.0  & 0    &0.2,1.8&1.0\nl
\enddata
\end{deluxetable}



\makeatletter
\def\jnl@aj{AJ}
\ifx\revtex@jnl\jnl@aj\let\tablebreak=\nl\fi
\makeatother


\begin{deluxetable}{cccccccc}
\tablewidth{0pc}
\tablecaption{Properties of the Central Region}
\tablehead{
\colhead{Model}           & \colhead{$n$}                 &
\colhead{$kT$}            & \colhead{$Z_{m}$}             &
\colhead{Model}           & \colhead{$n$}              &
\colhead{$kT$}            & \colhead{$Z_{m}$}          
\\
\colhead{}                &\colhead{($\rm \;cm^{-3}$)}&
\colhead{(keV)}           &\colhead{($Z_{\sun}$)}     &
\colhead{}                &\colhead{($\rm \;cm^{-3}$)}&
\colhead{(keV)}           &\colhead{($Z_{\sun}$)}     
}

\startdata
A1   & 0.2   & 0.8   & 2     & B4 & 0.4 & 1.1   & 1    \nl
A2   &\nodata&\nodata&\nodata& B5 & 10  & 1.2   & 1    \nl
A3   & 0.3   & 0.8   & 1     & B6 & 2   & 0.8   & 0.1  \nl
A4   & 0.4   & 0.9   & 1     &    &     &       &      \nl
A5   & 10    & 1.0   & 1     &    &     &       &      \nl
A6   & 0.7   & 0.6   & 0.1   &    &     &       &      \nl
\enddata
\end{deluxetable}

\end{document}